\documentclass[aps,prl,preprint,tightenlines,superscriptaddress,showpacs,byrevtex]{revtex4}
\usepackage{graphicx} % Include figure files
\usepackage{dcolumn}  % Align table columns on decimal point

\begin{document}

%\vspace*{-3\baselineskip}
%\resizebox{!}{3cm}{\includegraphics{belle.eps}}

\preprint{\vbox{ \hbox{   }
    \hbox{BELLE-CONF-0326}
    \hbox{EPS Parallel Session 10}
    \hbox{EPS-ID 553}
  }}

\title{\boldmath Evidence for $B^+\rightarrow\omega l^+ \nu$}

%%%% insert the authorlist here. BEFORE the abstract !!!
%%% Paper:    (fill in)
%%% Journal:  Summer 2003 conference proceedings (Physical Review format)
%%% Contacts: (fill in)
%%% Each author is included unless he/she chooses to opt out of ALL papers.
%%% ====================================================================
%%% Click the RELOAD button on your web browser to see the updated file.
%%% ====================================================================
%%% Use \input{author} to insert this material into your latex file.
%%%%% Force institutions to appear in alphabetical order when typeset.
\affiliation{Aomori University, Aomori}
\affiliation{Budker Institute of Nuclear Physics, Novosibirsk}
\affiliation{Chiba University, Chiba}
\affiliation{Chuo University, Tokyo}
\affiliation{University of Cincinnati, Cincinnati, Ohio 45221}
\affiliation{University of Frankfurt, Frankfurt}
\affiliation{Gyeongsang National University, Chinju}
\affiliation{University of Hawaii, Honolulu, Hawaii 96822}
\affiliation{High Energy Accelerator Research Organization (KEK), Tsukuba}
\affiliation{Hiroshima Institute of Technology, Hiroshima}
\affiliation{Institute of High Energy Physics, Chinese Academy of Sciences, Beijing}
\affiliation{Institute of High Energy Physics, Vienna}
\affiliation{Institute for Theoretical and Experimental Physics, Moscow}
\affiliation{J. Stefan Institute, Ljubljana}
\affiliation{Kanagawa University, Yokohama}
\affiliation{Korea University, Seoul}
\affiliation{Kyoto University, Kyoto}
\affiliation{Kyungpook National University, Taegu}
\affiliation{Institut de Physique des Hautes \'Energies, Universit\'e de Lausanne, Lausanne}
\affiliation{University of Ljubljana, Ljubljana}
\affiliation{University of Maribor, Maribor}
\affiliation{University of Melbourne, Victoria}
\affiliation{Nagoya University, Nagoya}
\affiliation{Nara Women's University, Nara}
\affiliation{National Kaohsiung Normal University, Kaohsiung}
\affiliation{National Lien-Ho Institute of Technology, Miao Li}
\affiliation{Department of Physics, National Taiwan University, Taipei}
\affiliation{H. Niewodniczanski Institute of Nuclear Physics, Krakow}
\affiliation{Nihon Dental College, Niigata}
\affiliation{Niigata University, Niigata}
\affiliation{Osaka City University, Osaka}
\affiliation{Osaka University, Osaka}
\affiliation{Panjab University, Chandigarh}
\affiliation{Peking University, Beijing}
\affiliation{Princeton University, Princeton, New Jersey 08545}
\affiliation{RIKEN BNL Research Center, Upton, New York 11973}
\affiliation{Saga University, Saga}
\affiliation{University of Science and Technology of China, Hefei}
\affiliation{Seoul National University, Seoul}
\affiliation{Sungkyunkwan University, Suwon}
\affiliation{University of Sydney, Sydney NSW}
\affiliation{Tata Institute of Fundamental Research, Bombay}
\affiliation{Toho University, Funabashi}
\affiliation{Tohoku Gakuin University, Tagajo}
\affiliation{Tohoku University, Sendai}
\affiliation{Department of Physics, University of Tokyo, Tokyo}
\affiliation{Tokyo Institute of Technology, Tokyo}
\affiliation{Tokyo Metropolitan University, Tokyo}
\affiliation{Tokyo University of Agriculture and Technology, Tokyo}
\affiliation{Toyama National College of Maritime Technology, Toyama}
\affiliation{University of Tsukuba, Tsukuba}
\affiliation{Utkal University, Bhubaneswer}
\affiliation{Virginia Polytechnic Institute and State University, Blacksburg, Virginia 24061}
\affiliation{Yokkaichi University, Yokkaichi}
\affiliation{Yonsei University, Seoul}
  \author{K.~Abe}\affiliation{High Energy Accelerator Research Organization (KEK), Tsukuba} % KEK
  \author{K.~Abe}\affiliation{Tohoku Gakuin University, Tagajo} % TohokuGakuin
  \author{N.~Abe}\affiliation{Tokyo Institute of Technology, Tokyo} % TIT
  \author{R.~Abe}\affiliation{Niigata University, Niigata} % Niigata
  \author{T.~Abe}\affiliation{High Energy Accelerator Research Organization (KEK), Tsukuba} % KEK
  \author{I.~Adachi}\affiliation{High Energy Accelerator Research Organization (KEK), Tsukuba} % KEK
  \author{Byoung~Sup~Ahn}\affiliation{Korea University, Seoul} % Korea
  \author{H.~Aihara}\affiliation{Department of Physics, University of Tokyo, Tokyo} % Tokyo
  \author{M.~Akatsu}\affiliation{Nagoya University, Nagoya} % Nagoya
  \author{M.~Asai}\affiliation{Hiroshima Institute of Technology, Hiroshima} % Hiroshima
% ...
% Lausanne
\collaboration{The Belle Collaboration}

\begin{abstract}
  We have searched for the decay~$B^+\rightarrow\omega l^+\nu$ in
78~fb$^{-1}$ of $\Upsilon(4S)$~data (85.0~million $B\bar B$ events)
accumulated with the Belle detector. The final state is fully
reconstructed using the $\omega$~decay into $\pi^+\pi^-\pi^0$ and
detector hermeticity to estimate the neutrino momentum. $155\pm
47$~signal events are found in the data, corresponding to a branching
fraction of $(1.3\pm 0.4\pm 0.2\pm 0.3)\cdot 10^{-4}$, where the first
two errors are statistical and systematic. The third error is due to
the estimated form-factor uncertainty. (This result is preliminary.)

\end{abstract}


\maketitle

%%%% keep the final version single-spaced
%\tighten

\section{Introduction}

The magnitude of $V_{ub}$ plays an important role in probing the
unitarity of the Cabbibo-Kobayashi-Maskawa (CKM)
matrix~\cite{ref:1}. At present, experimental constraints on
$|V_{ub}|$ come mainly from the analysis of the decay~$B\rightarrow
X_ul\nu$. Both inclusive approaches (sensitive
to all $X_ul\nu$~final states within a given region of phase space)
and the exclusive reconstruction of specific final states have been
attempted. For the latter, Belle has previously obtained preliminary
results for the decays $B\rightarrow\pi l\nu$ and
$B\rightarrow\rho l\nu$~\cite{ref:2}. This article presents
a study of the decay~$B^+\rightarrow\omega l^+\nu$, which has not been
observed so far~\cite{ref:3}.

Using a 78~fb$^{-1}$ dataset recorded on the $\Upsilon(4S)$~resonance
(85.0~million $B\bar B$ events), events with a single (undetected)
neutrino are selected and the neutrino momentum is inferred from
detector hermeticity. The neutrino candidate is combined with an
identified lepton (electron or muon) and an $\omega$ (reconstructed
through $\omega\rightarrow\pi^+\pi^-\pi^0$) and the final state is
reconstructed. The $B\rightarrow\omega l\nu$~yield and the remaining
background are determined by a binned maximum likelihood fit.

\section{Experimental procedure}

\subsection{KEKB and the Belle detector}

Belle is located at KEKB, an asymmetric $e^+e^-$~collider operating at
the center-of-mass energy of the $\Upsilon(4S)$~resonance. The
detector is described in detail elsewhere~\cite{ref:5}. It is a
large-solid-angle magnetic spectrometer that consists of a three-layer
silicon vertex detector (SVD), a 50-layer central drift chamber (CDC),
an array of aerogel threshold \v{C}erenkov counters (ACC), a
barrel-like arrangement of time-of-flight scintillation counters
(TOF), and an electromagnetic calorimeter comprised of CsI(Tl)
crystals (ECL) located inside a super-conducting solenoid coil that
provides a 1.5~T magnetic field.

The responses of the ECL, CDC ($dE/dx$) and ACC detectors are combined
to provide clean electron identification. Muons are identified in
the instrumented iron flux-return (KLM) located outside of the
coil. Charged hadron identification relies on the information from the
CDC, ACC and TOF~sub-detectors.

\subsection{Dataset}

The $\Upsilon(4S)$~dataset used for this study corresponds to an
integrated luminosity of 78.13~fb$^{-1}$ and contains $(85.0\pm
0.5)\cdot 10^{6}$~$B\bar B$~events. Continuum background is
subtracted using 8.83~fb$^{-1}$ of data taken below the
resonance.

A full detector simulation based on GEANT is applied to Monte Carlo
events. This analysis uses background Monte Carlo samples, equivalent
to about three times the integrated luminosity. Monte Carlo data
for the signal decay, $B^+\rightarrow\omega l^+\nu$, are generated
with three different form-factor models: ISGW2 (quark
model~\cite{ref:6}), UKQCD (quenched lattice QCD
calculation~\cite{ref:7}) and LCSR (light cone sum
rules~\cite{ref:8}). To model the cross-feed from other $B\rightarrow
X_ul\nu$~decays, Monte Carlo samples generated with the ISGW2 and the
De Fazio-Neubert model~\cite{ref:8b} are used.

\subsection{Neutrino reconstruction}

Events passing the hadronic selection are required to contain a single
lepton (electron or muon) with a c.m.\ momentum~$p^*_l$ greater than
1.3~GeV/$c$. In this momentum range, electrons (muons) are selected
with an efficiency of 92\% (89\%) and a pion fake rate of 0.25\%
(1.4\%).

The missing four-momentum is computed for selected events,
\begin{eqnarray}
  \vec p_{miss} & = & \vec p_{HER}+\vec p_{LER}-\sum_i\vec p_i~,
  \nonumber \\
  E_{miss} & = & E_{HER}+E_{LER}-\sum_i E_i~,
\end{eqnarray}
where the sum runs over all reconstructed charged tracks and
photons. The indices HER and LER refer to the high energy and the low
energy rings, respectively. To reject events in which the
missing momentum misrepresents the neutrino momentum, the following
selections are applied. Events with a large charge imbalance are
eliminated, $|Q_{tot}|<3e$, and the direction of the missing momentum is
required to lie within the ECL~acceptance,
$17^\circ<\theta_{miss}<150^\circ$. The missing mass squared,
$m^2_{miss}=E^2_{miss}-\vec p^2_{miss}$, is required to be consistent
with the neutrino hypothesis, $|m^2_{miss}|<3$~GeV$^2$/$c^4$.

After applying these cuts, the resolution in $p_{miss}$ is around
140~MeV/$c$ for generic $B\rightarrow
X_ul\nu$~events. As the energy resolution is worse
than the momentum resolution, the neutrino four-momentum is taken to
be $(p_{miss},\vec p_{miss})$. The efficiency of the combined event
selection and neutrino reconstruction cuts is about 17\% for
$B\rightarrow X_ul\nu$~events.

\subsection{Final state reconstruction}

Pairs of $\gamma$'s are combined to form
$\pi^0$~candidates ($E_\gamma>30$~MeV,
$120<m(\gamma\gamma)<150$~MeV/$c^2$). The
decay~$\omega\rightarrow\pi^+\pi^-\pi^0$ (branching ratio: $(89.1\pm
0.7)$\%~\cite{ref:9}) is reconstructed from all possible
combinations of one $\pi^0$ with two oppositely charged
tracks. Combinations with a charged track identified as a kaon are
rejected and the following selections are imposed:
$p^*_\omega>300$~MeV/$c$,
$703<m(\pi^+\pi^-\pi^0)<863$~MeV/$c^2$. Combinations located far
from the center of the Dalitz plot are removed by requiring the Dalitz
amplitude, $A\propto |\vec p_{\pi^+}\times \vec p_{\pi^-}|$, to be
larger than half of its maximum value.

The lepton in the event is combined with the $\omega$~candidate
and the neutrino, and the lepton momentum requirement is tightened,
$1.8<p^*_l<2.7$~GeV/$c$. To reject combinations inconsistent with
signal decay kinematics, the cut
$|\cos\theta_{BY}|<1.1$ is imposed,
\begin{equation}
  \cos\theta_{BY}=\frac{2E^*_B E^*_Y-m^2_B-m^2_Y}{2p^*_Bp^*_Y}~,
\end{equation}
where $E^*_B$, $p^*_B$ and $m_B$ are fixed to their nominal values,
$\sqrt{E_{HER}E_{LER}}$, $\sqrt{E^{*2}_B-m^2_B}$ and 5.279~GeV/$c^2$,
respectively. The variables~$E^*_Y$, $p^*_Y$ and $m_Y$ are the
measured c.m.\ energy, momentum and mass of the $Y=\omega+l$~system,
respectively. For well-reconstructed signal events, $\cos\theta_{BY}$
is the cosine of the angle between the $B$ and the $Y$~system
and lies between $-1$ and $+1$ while for background, a
significant fraction is outside this interval.

For each $\omega l\nu$~candidate, the beam-constrained mass $m_{bc}$
and $\Delta E$ are calculated ($E^*_{beam}=\sqrt{E_{HER}E_{LER}}$),
\begin{eqnarray}
  m_{bc} & = & \sqrt{(E^*_{beam})^2-|\vec p^*_\omega+\vec p^*_l+\vec
    p^*_\nu |^2}~, \nonumber \\
  \Delta E & = & E^*_{beam}-(E^*_\omega+E^*_l+E^*_\nu)~,
\end{eqnarray}
and candidates in the range $m_{bc}>5.23$~GeV/$c^2$ and
$|\Delta E|<1.08$~GeV are selected. On average, 2.5~combinations per
event satisfy all cuts and we choose the one with the largest
$\omega$~momentum in the c.m.\ frame. This choice is correct in
77\% of the cases.

\subsection{Continuum suppression}

The background from continuum~$e^+e^-\rightarrow q\bar q$~events,
$q=u,d,s,c$, is suppressed using three variables that exploit the fact
that, in the $\Upsilon(4S)$ frame, the two $B$~mesons are produced
nearly at rest and that therefore $B\bar B$~events are nearly spherical 
while continuum events have a more jet-like topology. These variables
are:
\begin{itemize}
  \item The ratio~$R_2$ of the second to the zeroth Fox-Wolfram
    moment~\cite{ref:10}. This ratio tends to be close to zero (unity)
    for spherical (jet-like) events. (The cut $R_2<0.4$ is imposed
    at the event selection level.)
  \item The cosine of $\theta_{thrust}$, where $\theta_{thrust}$ is
    the angle between the thrust axis of the $\omega l$~system and the
    thrust axis of the rest of the event.
  \item A Fisher discriminant that selects events with an uniform energy
    distribution around the lepton direction~\cite{ref:11}. The input
    variables are the charged and neutral energy in nine cones of
    equal solid angle around the lepton momentum axis.
\end{itemize}

A cut on the likelihood ratio combining the three variables is
imposed. This selection is 56\%~efficient for $B^+\rightarrow\omega
l^+\nu$ events while it eliminates 92\% of the continuum background
remaining after the $R_2<0.4$~cut.

\subsection{The fit}

The signal yield and the remaining background are determined by a
binned maximum likelihood fit in the $\Delta E$ vs.\
$m(\pi^+\pi^-\pi^0)$ plane taking into account the finite Monte Carlo
statistics~\cite{ref:12}. 9 bins in $\Delta E$ (bin width: 240~MeV)
and 8 bins in $m(\pi^+\pi^-\pi^0)$ (bin width: 20~MeV/$c^2$) are
used. This fit is performed simultaneously in three bins of lepton
momentum, $1.8<p^*_l<2.1$~GeV/$c$, $2.1<p^*_l<2.4$~GeV/$c$ and
$2.4<p^*_l<2.7$~GeV/$c$.

\begin{figure}
  \begin{center}
    \includegraphics[width=13.8cm]{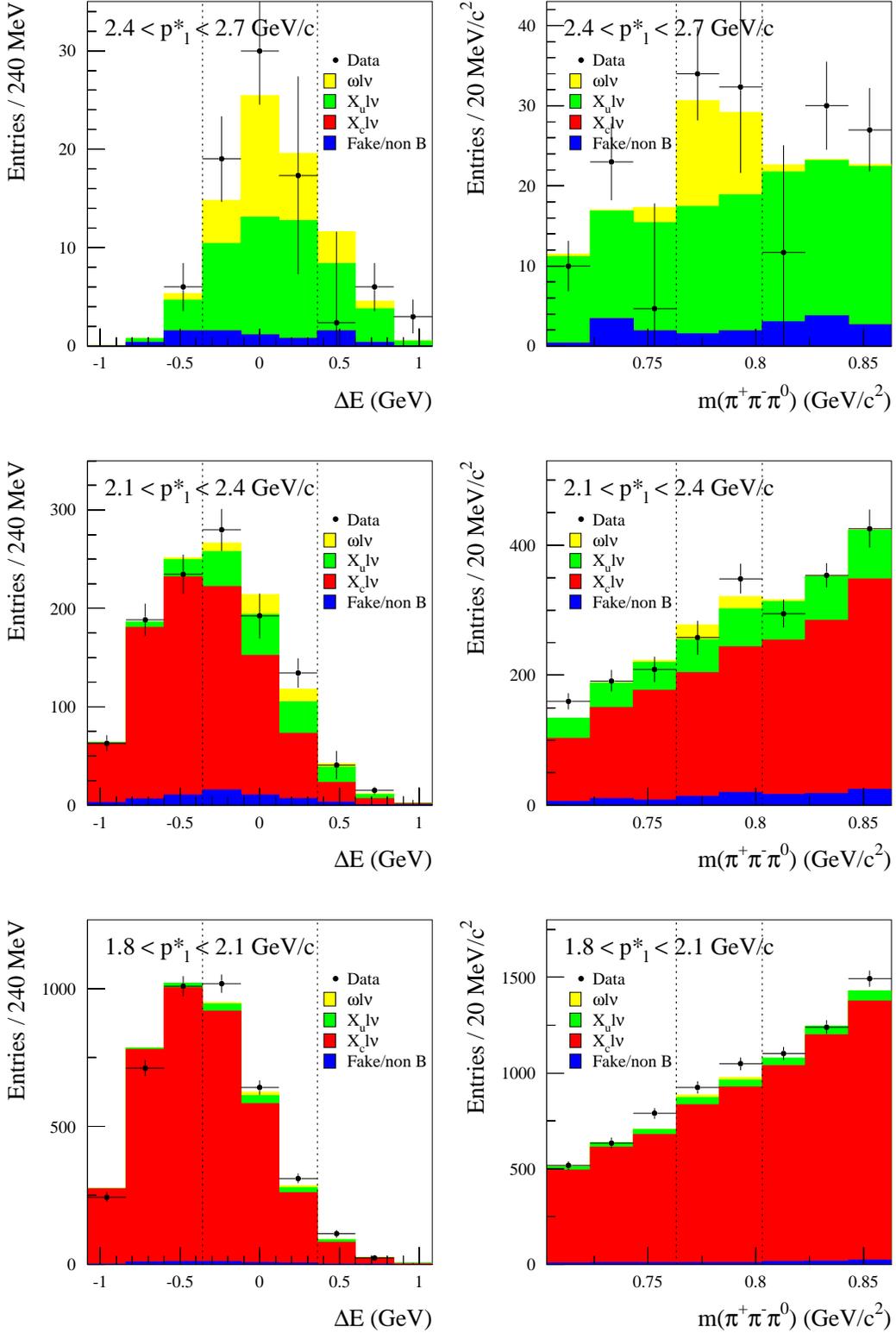}
  \end{center}
  \caption{The result of the fit assuming ISGW2~form-factors for
    $B^+\rightarrow\omega l^+\nu$. The $\Delta E$ and
    $m(\pi^+\pi^-\pi^0)$~distributions in each $p^*_l$~bin are shown.
    The data points are continuum subtracted
    on-resonance data, the histograms are the components of the fit,
    as described in the text.} \label{fig:1}
\end{figure}

Five components are fitted to the data: the $B^+\rightarrow\omega l^+\nu$
signal, $B\rightarrow X_ul\nu$ background, $B\rightarrow X_cl\nu$
background, fake and non-$B$~decay lepton background and
continuum. The shapes of the first four components are determined
from simulation, while the shape of the continuum component is given by
the off-resonance data. The normalizations of the
$B^+\rightarrow\omega l^+\nu$, the $B\rightarrow X_ul\nu$ and the
$B\rightarrow X_cl\nu$~components are floated in the fit, while the other
components are fixed (Table~\ref{tab:1}, Fig.~\ref{fig:1}).
\begin{table}
  \begin{center}
    \begin{tabular}{|c|c|c|c|}
      \hline
      \rule[-1.3ex]{0pt}{4ex} & $1.8<p^*_l<2.1$~GeV/$c$ &
      $2.1<p^*_l<2.4$~GeV/$c$ & $2.4<p^*_l<2.7$~GeV/$c$\\
      \hline \hline
      \rule[-1.3ex]{0pt}{4ex}data & 1990 & 667 & 75\\
      \hline \hline
      \rule[-1.3ex]{0pt}{4ex}$B^+\rightarrow\omega l^+\nu$ & $41\pm
      13$ & $68\pm 21$ & $35\pm 11$\\
      \rule[-1.3ex]{0pt}{4ex}$B\rightarrow X_ul\nu$ & $61\pm 28$ &
      $82\pm 28$ & $21\pm 5$\\
      \rule[-1.3ex]{0pt}{4ex}$B\rightarrow X_cl\nu$ & $1743\pm 36$ &
      $415\pm 14$ & 0\\
      \rule[-1.3ex]{0pt}{4ex}fake, non $B$ & $19\pm 3$ & $33\pm 4$ &
      $3\pm 1$\\
      \rule[-1.3ex]{0pt}{4ex}continuum & $17\pm 12$ & $61\pm 23$ &
      $9\pm 9$\\
      \hline
      \rule[-1.3ex]{0pt}{4ex}sum & $1881\pm 49$ & $659\pm 44$ & $68\pm
      15$\\
      \hline
    \end{tabular}
  \end{center}
  \caption{The result of the fit assuming ISGW2~form-factors for
    $B^+\rightarrow\omega l^+\nu$. For each $p^*_l$~bin,
    the number of events in the signal window,
    $763<m(\pi^+\pi^-\pi^0)<803$~MeV/$c^2$ and $|\Delta E|<360$~MeV,
    are shown for the data and the different components of the fit.}
    \label{tab:1}
\end{table}

\section{Result and systematic uncertainty}

The fit is repeated for each of the three form-factor models available
for $B^+\rightarrow\omega l^+\nu$. For each model, the signal yield,
$N(B^+\rightarrow\omega l^+\nu)$, is determined and the branching
ratio, ${\mathcal B}(B^+\rightarrow\omega l^+\nu)$, is calculated
accordingly using the relation (Table~\ref{tab:2})
\begin{equation}
  N(B^+\rightarrow\omega l^+\nu)=N(B^+)\times{\mathcal
    B}(B^+\rightarrow\omega l^+\nu)\times {\mathcal
    B}(\omega\rightarrow\pi^+\pi^-\pi^0)\times(\epsilon_e+\epsilon_\mu)~.
\end{equation}
$N(B^+)$ is the total number of charged $B$~mesons in the
data, ${\mathcal B}(\omega\rightarrow\pi^+\pi^-\pi^0)=(89.1\pm
0.7)\%$~\cite{ref:9} and $\epsilon_e$ ($\epsilon_\mu$) is the
model-dependent selection efficiency for $\omega e\nu$
($\omega\mu\nu$) candidates.
Averaging over the three models (giving equal weight to
each), a branching fraction of $(1.29\pm
0.39)\cdot 10^{-4}$ is obtained. The spread around this average value,
which amounts to $0.28\cdot 10^{-4}$, is used as an estimate of the
form-factor model uncertainty.
\begin{table}
  \begin{center}
    \begin{tabular}{|c|c|c|c|}
      \hline
      \rule[-1.3ex]{0pt}{4ex}form-factor model & signal yield &
      ${\mathcal B}(B^+\rightarrow\omega l^+\nu)$ & $\chi^2/\mathrm{ndf}$\\
      \hline \hline
      \rule[-1.3ex]{0pt}{4ex}ISGW2 & $144\pm 44$ & $(1.00\pm
      0.31)\cdot 10^{-4}$ & 1.05\\
      \rule[-1.3ex]{0pt}{4ex}UKQCD & $145\pm 44$ & $(1.20\pm
      0.37)\cdot 10^{-4}$ & 1.08\\
      \rule[-1.3ex]{0pt}{4ex}LCSR & $176\pm 52$ & $(1.67\pm
      0.50)\cdot 10^{-4}$ & 1.04\\
      \hline
      \rule[-1.3ex]{0pt}{4ex}average & $155\pm 47\pm 15$ & $(1.29\pm
      0.39\pm 0.28)\cdot 10^{-4}$ & \\
      \hline
    \end{tabular}
  \end{center}
  \caption{The yield in the signal window,
    $763<m(\pi^+\pi^-\pi^0)<803$~MeV/$c^2$ and $|\Delta E|<360$~MeV,
    the corresponding branching fraction and the goodness of fit
    (estimated by the $\chi^2$ divided by the
    number of degrees of freedom). For each fit (using a given
    form-factor model), the error quoted on the signal yield and the
    branching fraction is statistical only. For the average, the
    first error is statistical and the second is the spread around the
    central value.} \label{tab:2}
\end{table}

The quadratic sum of the experimental systematics (listed in
Table~\ref{tab:3}) is $0.21\cdot 10^{-4}$ or 16.4\% of the branching
fraction. The largest contribution is the uncertainty in the
$X_ul\nu$~cross-feed. It is estimated by separately varying the
fraction of $B\rightarrow\pi l\nu$ and $B\rightarrow\rho l\nu$~decays
(that are expected to dominate in the high $p^*_l$~region) within their
respective experimental uncertainties. For the cross-feed from other
$B\rightarrow X_ul\nu$~decays, the fit is repeated modeling this
component once with the ISGW2 (fully resonant $B\rightarrow X_ul\nu$)
and once with the De Fazio-Neubert model (fully non-resonant
$B\rightarrow X_ul\nu$). Half of the difference between these two
extreme cases is assigned as a systematic uncertainty.
\begin{table}
  \begin{center}
    \begin{tabular}{|c|c|c|c|}
      \hline
      \rule[-1.3ex]{0pt}{4ex} & value & $\quad \Delta{\mathcal B}/{\mathcal
        B}\quad $ & $\quad$Ref.$\quad$\\
      \hline \hline
      \rule[-1.3ex]{0pt}{4ex}$B^0\rightarrow\pi^-l^+\nu$ &
      $(1.8\pm 0.6)\cdot 10^{-4}$ & 2.2\% & \cite{ref:9}\\
      \rule[-1.3ex]{0pt}{4ex}$B^0\rightarrow\rho^-l^+\nu$ &
      $(2.6^{+0.6}_{-0.7})\cdot 10^{-4}$ & 12.7\% & \cite{ref:9}\\
      \rule[-1.3ex]{0pt}{4ex}other $B\rightarrow X_ul\nu$ &
      & 1.4\% & \\
      \hline
      \rule[-1.3ex]{0pt}{4ex}$X_ul\nu$ cross-feed (sum) & & 13.0\% & \\
      \hline \hline
      \rule[-1.3ex]{0pt}{4ex}neutrino reconstruction & & 4\% & \\
      \rule[-1.3ex]{0pt}{4ex}charged track finding ($l,\pi^+,\pi^-$) & &
      3\% & \\
      \rule[-1.3ex]{0pt}{4ex}cluster finding ($\pi^0$) & & 4\% & \\
      \hline
      \rule[-1.3ex]{0pt}{4ex}(sum) & & 9\% & \\
      \hline \hline
      \rule[-1.3ex]{0pt}{4ex}$X_cl\nu$ cross-feed & & 2.8\% & \\
      \rule[-1.3ex]{0pt}{4ex}lepton identification & & 3.0\% & \\
      \rule[-1.3ex]{0pt}{4ex}number of $B\bar B$ & $(85.0\pm 0.5)\cdot
      10^6$ & 0.6\% & \\
      \rule[-1.3ex]{0pt}{4ex}${\mathcal
        B}(\omega\rightarrow\pi^+\pi^-\pi^0)$ & $(89.1\pm 0.7)\%$ &
      0.8\% & \cite{ref:9}\\
      \hline \hline
      total systematic uncertainty & & 16.4\% & \\
      \hline
    \end{tabular}
  \end{center}
  \caption{Contributions to the systematic
      uncertainty. The size of each contribution is given as
      percentage of the branching ratio. The different components are
      discussed in the text.} \label{tab:3}
\end{table}

The next largest component is the uncertainty in the neutrino
reconstruction, track finding and cluster finding efficiency. Other
contributions taken into account in the calculation of the systematic
error are: $X_cl\nu$~cross-feed (estimated by varying the fraction of
$B\rightarrow D^*l\nu$ in the $X_c l\nu$~component), lepton
identification, the number of $B\bar B$~events and the uncertainty in
the $\omega\rightarrow\pi^+\pi^-\pi^0$ branching fraction.

\section{Conclusion}

We have studied the decay~$B^+\rightarrow\omega l^+\nu$ using
78~fb$^{-1}$ of $\Upsilon(4S)$ data (85.0~million $B\bar B$
events). The final state was reconstructed using
the $\omega$~decay into $\pi^+\pi^-\pi^0$ and detector hermeticity to
infer the neutrino momentum. The signal yield and the remaining
background were estimated by a binned maximum-likelihood
fit. Repeating the fit for three different $B^+\rightarrow\omega
l^+\nu$ form-factor models and averaging the result, $155\pm 47$
signal events are found, corresponding to a branching fraction of
$(1.3\pm 0.4\pm 0.2\pm 0.3)\cdot 10^{-4}$, where the
errors are statistical, systematic and estimated form-factor
uncertainty, respectively. (This result is preliminary.)

\section{Acknowledgements}
% Please paste this acknowledgement into your latex file.
%***** Acknowledgments *****
We wish to thank the KEKB accelerator group for the excellent
operation of the KEKB accelerator.
We acknowledge support from the Ministry of Education,
Culture, Sports, Science, and Technology of Japan
and the Japan Society for the Promotion of Science;
the Australian Research Council
and the Australian Department of Education, Science and Training;
the National Science Foundation of China under contract No.~10175071;
the Department of Science and Technology of India;
the BK21 program of the Ministry of Education of Korea
and the CHEP SRC program of the Korea Science and Engineering Foundation;
the Polish State Committee for Scientific Research
under contract No.~2P03B 01324;
the Ministry of Science and Technology of the Russian Federation;
the Ministry of Education, Science and Sport of the Republic of Slovenia;
the National Science Council and the Ministry of Education of Taiwan;
and the U.S.\ Department of Energy.


\begin{thebibliography}{99}
\bibitem{ref:3} Throughout this article, the inclusion of the charge
  conjugate mode decay is implied.
\end{thebibliography}
\end{document}